\documentclass[aps,prb,twocolumn,floatfix,superscriptaddress]{revtex4-1}

\usepackage[utf8]{inputenc}
\usepackage{t1enc}
\usepackage{graphicx}
\usepackage{epstopdf}
\usepackage{float}
\usepackage{color}
\usepackage{xcolor,colortbl}
\usepackage{amsmath}
\usepackage[caption=false,subrefformat=parens]{subfig}
\usepackage{comment}
\usepackage{times}

\begin{document}
\title{\emph{Ab initio} theory of $\text{N}_{2}\text{V}$ defect as quantum memory in diamond}
\author{P\'eter Udvarhelyi}
\affiliation{Loránd Eötvös University, Pázmány Péter sétány 1/A, H-1117 Budapest, Hungary}
\affiliation{Institute for Solid State Physics and Optics, Wigner Research Centre for Physics, Budapest, Hungary}
\author{Gerg\H{o} Thiering}
\affiliation{Institute for Solid State Physics and Optics, Wigner Research Centre for Physics, Budapest, Hungary}
\affiliation{Budapest University of Technology and Economics, Budafoki út 8., H-1111 Budapest, Hungary}
\author{Elisa Londero}
\affiliation{Institute for Solid State Physics and Optics, Wigner Research Centre for Physics, Budapest, Hungary}
\author{Adam Gali}
\email{gali.adam@wigner.mta.hu}
\affiliation{Institute for Solid State Physics and Optics, Wigner Research Centre for Physics, Budapest, Hungary}
\affiliation{Budapest University of Technology and Economics, Budafoki út 8., H-1111 Budapest, Hungary}
\date{\today}

\begin{abstract}
$\text{N}_{2}\text{V}$ defect in diamond is characterized by means of \emph{ab initio} methods relying on density functional theory calculated parameters of a Hubbard model Hamiltonian. It is shown that this approach appropriately describes the energy levels of correlated excited states induced by this defect. By determining its critical magneto-optical parameters, we propose to realize a long-living quantum memory by $\text{N}_{2}\text{V}$ defect in diamond.
\end{abstract}

\maketitle

\section{Introduction}

\begin{figure*}
\centering
\subfloat[\label{N2Vgeom}]{
\includegraphics[scale=0.06]{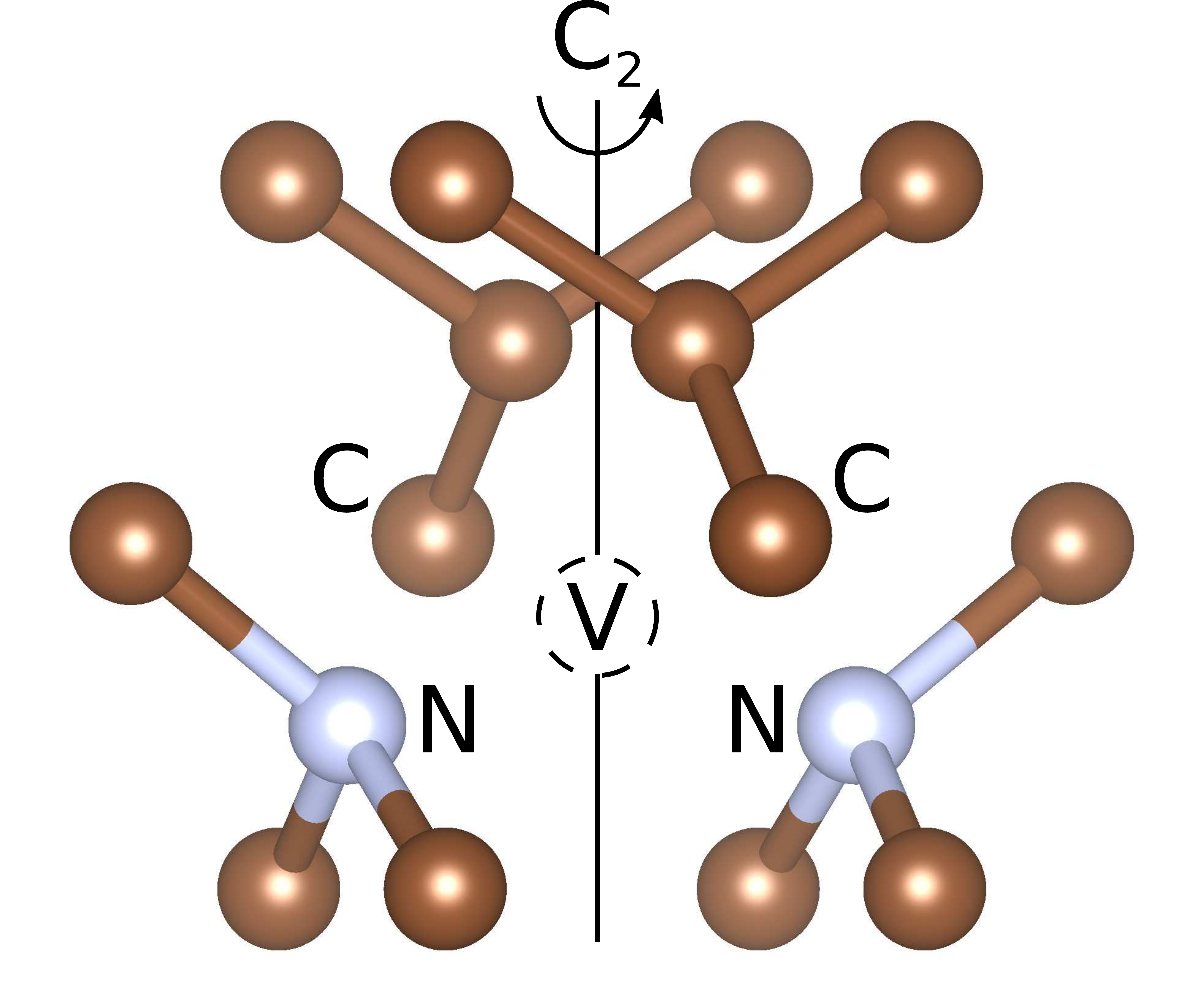}
}
\hfill
\subfloat[\label{N2Vstates}]{
\includegraphics[scale=0.6]{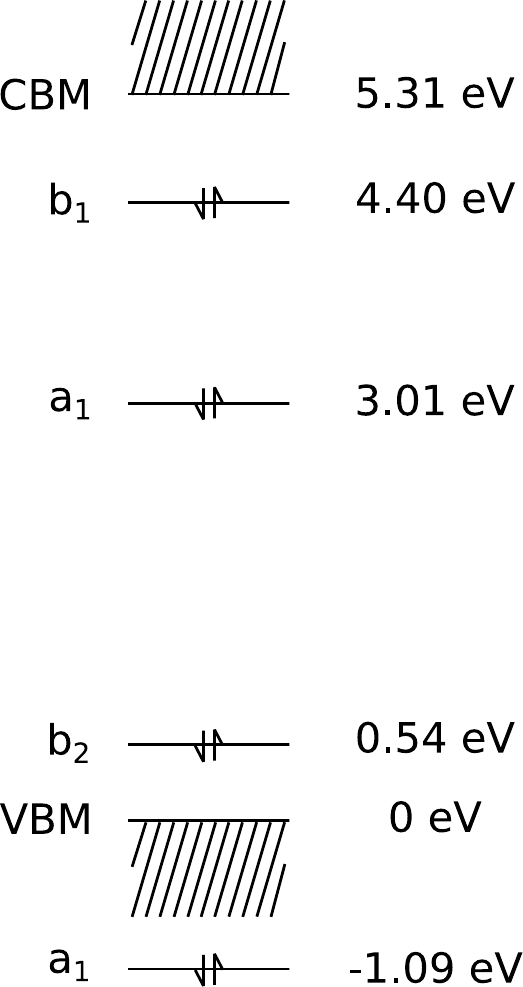}
}
\hfill
\subfloat[\label{ODMR}]{
\includegraphics[scale=0.7]{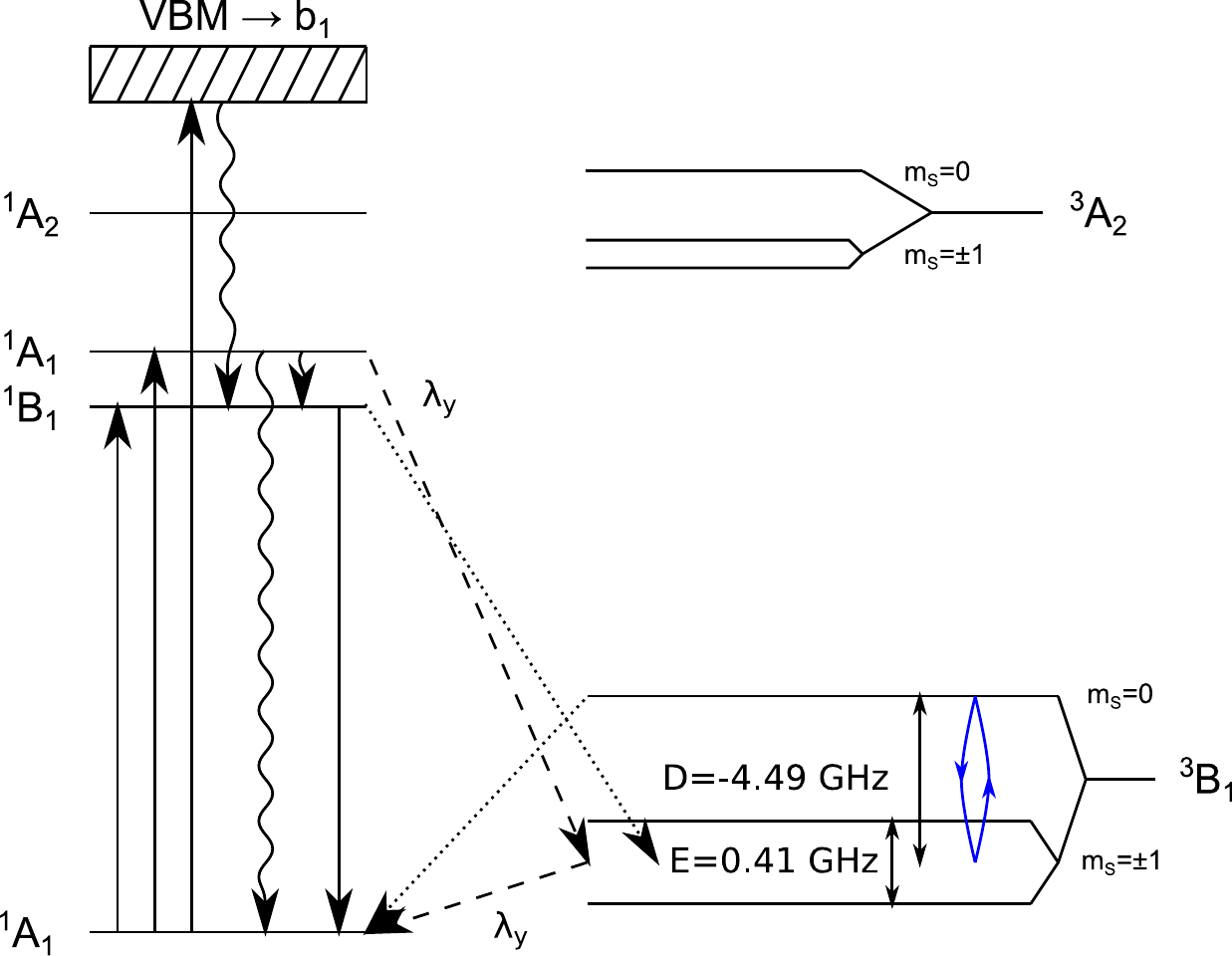}
}
\caption{\protect\subref{N2Vgeom} Geometry of $\text{N}_{2}\text{V}$ defect in diamond. \protect\subref{N2Vstates} Representation of defect levels of double negatively charged $\text{N}_{2}\text{V}$ with closed-shell orbitals relative to the valence band maximum (VBM). \protect\subref{ODMR} Analysis of ODMR contrast of the neutral $\text{N}_{2}\text{V}$. Straight line illustrates radiative decay, dashed and dotted lines represent ISC with first order and second-order spin-orbit couplings ($\lambda$), respectively. Curved lines and blue arrows indicate phonon and microwave transition, respectively. The calculated zero field splitting parameters are given for $^{3}B_{1}$ state at 0~K temperature.}
\label{fig:1}
\end{figure*}

Paramagnetic point defects in diamond are candidates for quantum bit and quantum information applications. Various defects were identified as optically active color centers\cite{diamond}, most notable is the nitrogen-vacancy (NV)~\cite{NVvibronic}. NV is formed in diamond by trapping of a mobile vacancy by the substitutional nitrogen atom. In nitrogen rich diamond aggregation of subtitutional nitrogen atoms may occur. The neighbor substitutional nitrogen pair is an example for such aggregation (A-aggregate). Similar to NV defect, $\text{N}_{2}\text{V}$ defect is formed by an A-aggregate trapping a mobile vacancy\cite{Mainwood}. Uniaxial stress measurements already established N-V-N structure of the $\text{N}_{2}\text{V}$ defect with $\text{C}_{2\text{V}}$ symmetry [see Fig.~\ref{fig:1}\subref{N2Vgeom}] \cite{aggregate}. The effective one-electron picture of the defect is described by Lowther\cite{lowther}. The defect in its neutral charge state was assigned to H3 color center\cite{jones1994} with zero phonon line (ZPL) of $2.463$~eV, where this signal was associated with the optical transition between $^1A_1$ ground and $^1B_1$ excited state\cite{Davies245}. The H3 photoluminescence (PL) center has a PL lifetime of $17.5$~ns and outstanding $0.95$ quantum yield\cite{crossfield1974} that makes the defect a stable single photon source\cite{single}. An optically inactive $^1A_1$ state with absorption line at $2.479$~eV was revealed under uniaxial stress measurements\cite{Davies245}. Furthermore, H13 absorption band with ZPL at $3.364$~eV was also observed and interpreted as transition to higher excited states\cite{H13}. The H3 center shows delayed luminescence, that was interpreted as a reversible transition from the $^1B_1$ excited state to metastable triplet states, with radiative decay time in the order of tens of milliseconds\cite{pereira}. Its paramagnetic metastable triplet state was investigated by electron spin resonance (ESR) spectroscopy, called W26 center, under illumination at room temperature. The experimental zero-field splitting (ZFS) tensor principal values are $D_{xx}=1.43$~GHz and $D_{zz}=-2.63$~GHz. The measured hyperfine parameters of $^{14}\text{N}$ nuclei are $A_{\perp}=10.2$~MHz and $A_{\parallel}=21.5$~MHz\cite{vanWyk}.

Our study is motivated by the success of optically detected magnetic resonance (ODMR) applications of single NV defect\cite{gruber1997} and the readout and control of single nuclear spin with ST1 defect in diamond\cite{Lee2013}. The latter employs optical pumping to metastable triplet state where initialization of the nuclear spin is achieved by spin polarization transfer exploiting the hyperfine level anticrossing (LAC). As the metastable triplet state relaxes to the singlet groundstate, the nuclear spin coherence time is not reduced by the persistent electron spin. As H3 center exhibits singlet groundstate and optically accessible metastable triplet state we wished to explore the properties of $\text{N}_{2}\text{V}$ defect for quantum memory applications.

To this end, we characterize this defect in diamond by means of advanced density functional theory (DFT) calculations. In this paper, we demonstrate that optical spin polarization of the triplet state and spin polarization transfer to the existing nuclear spins is principally feasible, i.e., a long living quantum memory may be realized with $\text{N}_{2}\text{V}$ defect. Their magneto optical parameters is determined by means of DFT calculations that go beyond the conventional Kohn-Sham DFT methods. By combining von~Barth theory\cite{vonBarth} and Hubbard model we developed an \emph{ab initio} method to calculate the energy of highly correlated multiplets with using only Kohn-Sham DFT wavefunction and energies, and we apply this to the neutral $\text{N}_{2}\text{V}$ defect.

We organized our paper as follows. In the next section (Sec.~\ref{sec:compmeth}), details about the computational method are given including test results on the negatively charged $\text{N}_{2}\text{V}$ defect. We focus then on the proposed quantum memory application of the neutral $\text{N}_{2}\text{V}$ defect in Sec.~\ref{sec:Qmem} that is the main topic of our paper. In Sec.~\ref{sec:Hubbard} we describe the Hubbard Hamiltonian analysis of the electronic structure of the neutral $\text{N}_{2}\text{V}$ defect. We report the calculated \emph{ab initio} magneto-optical parameters in Sec.~\ref{sec:results} that are taken in the quantum memory discussion in Sec.~\ref{sec:Qmem}.

\section{Computational method}
\label{sec:compmeth}
   
We carried out DFT calculations for electronic structure calculation and geometry optimization within spinpolarized HSE06 hybrid functional\cite{HSE} using the plane wave and projector augmented-wave (PAW) formalism based Vienna Ab initio Simulation Package (VASP)\cite{paw, VASP1,VASP2,VASP3,VASP4}. The model of $\text{N}_{2}\text{V}$ in bulk diamond was constructed using a 512-atom diamond simple cubic supercell within the $\Gamma$-point approximation. The $\Gamma$-point approximation simplifies the process of ensuring
the proper symmetry of the Kohn-Sham wave functions which is advantageous in our method. The $\Gamma$-point sampling of the Brillouin-zone has proven sufficient for various defects in diamond for 512-atom supercell \cite{deak2014, Kaviani}. This implies that the employed parameters provide sufficiently converged results for $\text{N}_{2}\text{V}$. Most of the calculations were performed with $370~\text{eV}$ plane wave cutoff energy that is sufficient for electronic structure of nitrogen-vacancy type defects calculations \cite{deak2014}. Hyperfine interaction parameters were obtained with core correction included\cite{szasz} with an increased cutoff energy of $500~\text{eV}$. Zero field splitting parameters were calculated with a home built code with the use of the same parameters and methods that are given in Ref.~\onlinecite{Viktor2014}.

HSE06 calculations provide excellent results for the negatively charged $\text{N}_{2}\text{V}$ defect which has a spin doublet with no high correlation between the electron states. To demonstrate this, we compare the experimental hyperfine constants \cite{Green2015} with our HSE06 DFT calculations in Table~\ref{sup:hyperfine2}, and we found excellent agreement. This supports to apply HSE06 functional for orbitals that are not highly correlated. However, we found that the neutral $\text{N}_{2}\text{V}$ is very challenging for Kohn-Sham DFT functionals because of the highly correlated open-shell orbitals. We present a method in Sec.~\ref{sec:Hubbard} that properly calculate these states that involves a Hubbard model Hamiltonian. Our method can be useful in the study of other quantum bits with highly correlated electronic states.
\begin{table}[H]
\caption{HSE06 DFT calculated and experimental (in parentheses) hyperfine principal values for the first neighbor nitrogen and first carbon atoms around the vacancy of the negatively charged $\text{N}_{2}\text{V}$ defect. The experimental data are taken from Ref.~\onlinecite{Green2015}.}
\label{sup:hyperfine2}
\begin{ruledtabular}
\begin{tabular}{lccc}
atom &$A_{xx}$ (MHz)&$A_{yy}$ (MHz)&$A_{zz}$ (MHz)\\
$^{15}\text{N}$&$4.0 (3.47)$&$4.5 (4.09)$&$5.0 (4.51)$\\
$^{13}\text{C}$&$190.8 (202.3)$&$191.6 (202.3)$&$314.3 (317.5)$\\ 
\end{tabular}
\end{ruledtabular}
\end{table}

\section{Proposed quantum memory application of the $\text{N}_{2}\text{V}$ defect}
\label{sec:Qmem}

By using HSE06 we found in an earlier study\cite{deak2014} that the $(+|0)$ and $(0|-)$ charge transition levels of $\text{N}_{2}\text{V}$ defect are at $E_\text{C}-4.8$~eV and $E_\text{V}+3.3$~eV, respectively, where $E_\text{C}$ and $E_\text{V}$ is the conduction and valence band edge, respectively. This explains the stability of its neutral charge state at various doping concentrations. Furthermore, its H13 absorption band can be associated with the transitions from the valence band to the empty in-gap defect level of the neutral defect. The method to calculate the lower energy states and electronic structure is given in the next Sections. Our basic results are summarized in Fig.~\ref{fig:1} that shows the optically induced electron spinpolarization process. We find two optically active excited states ($^{1}B_{1}$ and $^{1}A_{1}$ with small energy gap) and an optically forbidden dark state ($^{1}A_{2}$). After the excitation to the phonon sideband of the optically allowed singlet excited states, it can relax to the vibronic groundstate of $^{1}B_1$, and then back to the $^{1}A_{1}$ electronic groundstate with emitting a photon. Alternatively, intersystem crossing (ISC) from $^{1}B_{1}$ to $^{3}B_{1}$ may take place too as a second order process, where mixing of the excited $^{1}B_{1}$ and $^{1}A_{1}$ states caused by $B_{1}$-type phonons makes the intersystem crossing feasible via spin-orbit interaction. At elevated temperatures, the $^{1}A_{1}$ excited state may be thermally occupied (experimental gap is $16$~meV in Ref.~\onlinecite{Davies245}) and then a first order ISC to $^{3}B_{1}$ can occur. This is a \emph{spin selective} transition to $m_{S}=\pm1$ as only $\lambda_y$ of $B_{1}$ symmetry can couple these states, where $\lambda_y$ is the $y$-component of the spin-orbit coupling. We note that the spin sublevels of the triplet split even at zero magnetic field that is caused by the electron spin - electron spin dipolar interaction (zero-field splitting) because of 
the low-symmetry crystal field. Thus, the spin selective ISC will indeed populate only the $m_{S}=\pm1$ states. From this metastable triplet, an ISC can occur to the singlet groundstate. Again, the transition from the triplet $m_{S}=\pm1$ and $m_{S}=0$ substates to the singlet $^1A_{1}$ groundstate is a first order and a second order process, respectively, because of the selection rules. The singlet-triplet ISC is expected to be significantly slower than the rate of the radiative decay, because the large gap between the excited state singlet and the metastable triplet according to our \emph{ab initio} result. This is in contrast to the interpretation of an experiment \cite{pereira}, that we will discuss below. This \emph{ab initio} result may explain the large quantum yield of the defect. In the optical cycle, ODMR contrast can be achieved by microwave excitation, owing to the lifetime differences of first and second order transitions from the different triplet substates to the singlet groundstate. 

The system shows characteristics that makes it a promising candidate for quantum memory applications. Electron spinpolarization can be achieved in the ODMR cycle by populating the $m_{S}=\pm1$ sublevel of the metastable triplet. The $^{14}\text{N}$ (or $^{15}\text{N}$ nuclei of the defect and $^{13}\text{C}$ nuclei in their vicinity are candidates for quantum memory. The calculated hyperfine constants are listed for these nuclear spins in Sec.~\ref{sec:results}.  Spinpolarization transfer between electron and nearby nuclear spins can be realized at LAC condition by optical pumping of the defect (see detailed analysis of this process in Refs.~\cite{Ivady2015, Ivady2016}). LAC condition can be realized by a constant external magnetic field which is perfectly aligned with the symmetry axis of the defect and its magnitude equals the zero-field-splitting \cite{Jacques2009, Lee2013}. 
After the nuclear spin was set, the electron will naturally decay to the singlet groundstate that does not decohere the nuclear spin.

\section{Hubbard model of the electronic states}
\label{sec:Hubbard}

\begin{figure}
\begin{minipage}{.7\linewidth}\hfill
\subfloat[\label{a1n}]{
\includegraphics[scale=0.08]{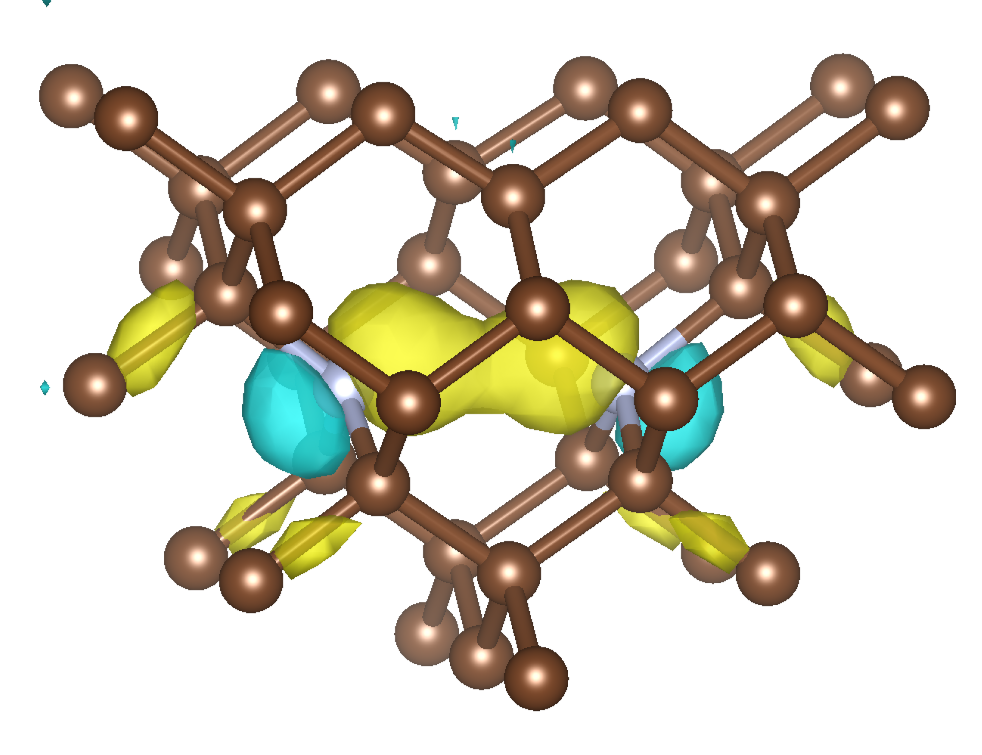}
}\hfill
\subfloat[\label{b2n}]{
\includegraphics[scale=0.08]{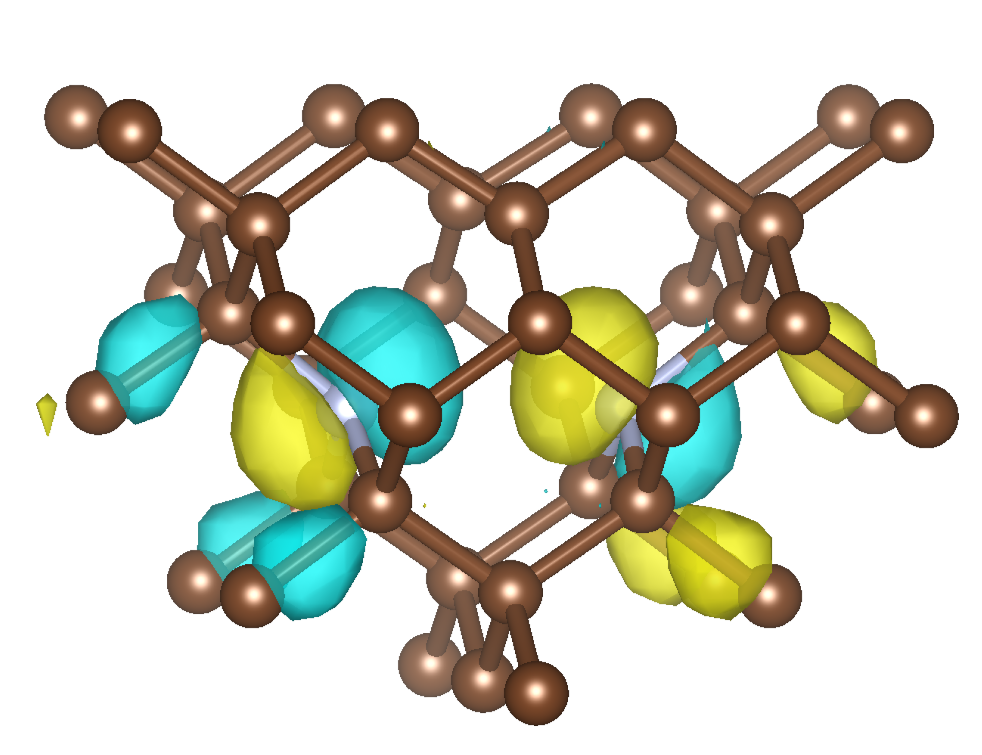}
}\hfill
\subfloat[\label{a1c}]{
\includegraphics[scale=0.08]{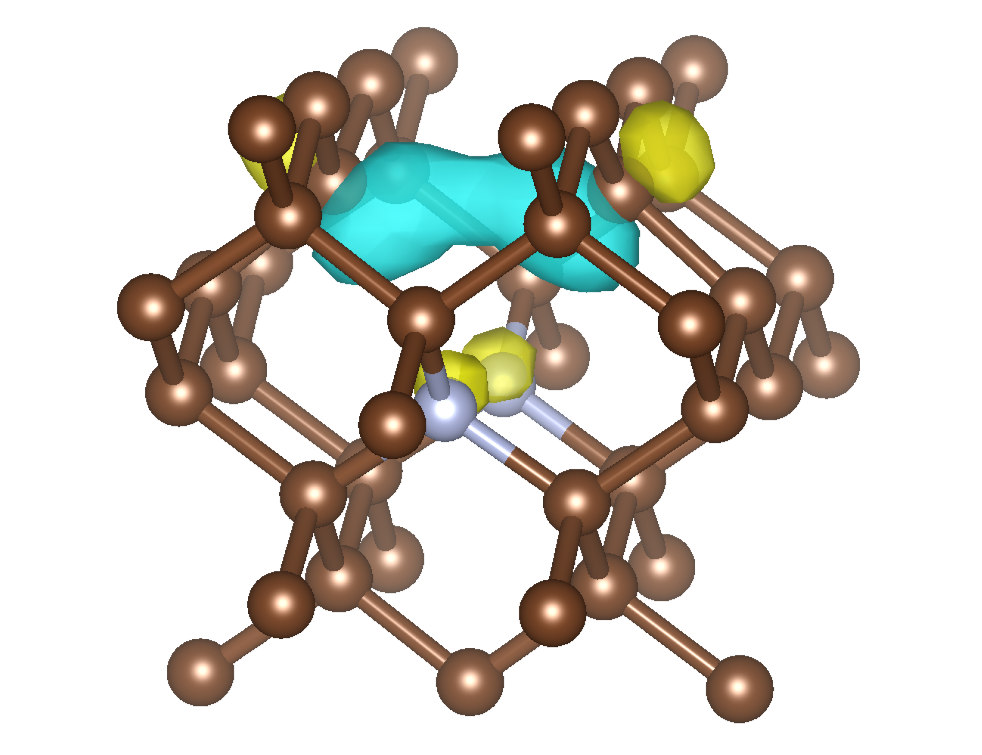}
}\hfill
\subfloat[\label{b1c}]{
\includegraphics[scale=0.08]{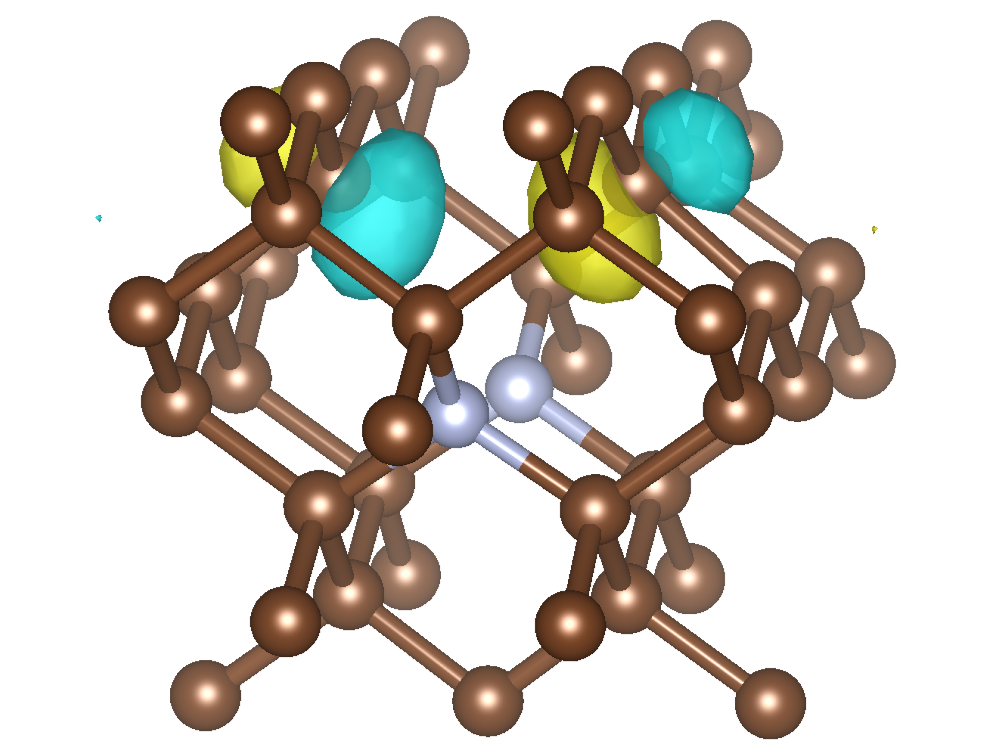}
}\hfill
\end{minipage}
\begin{minipage}{.2\linewidth}\hfill
\subfloat[\label{sup:states}]{
\includegraphics[scale=0.6]{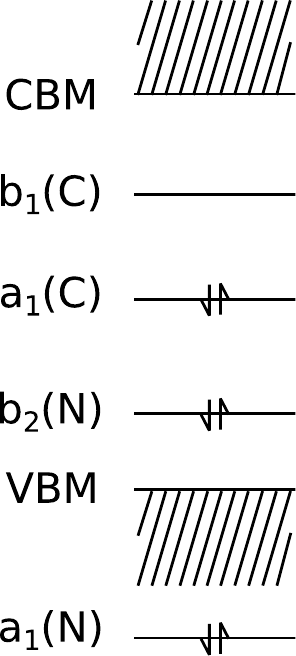}
}\hfill
\end{minipage}
\caption{Visualisation of geometric structure and defect wavefunctions of $\text{N}_{2}\text{V}$. Brown and blue balls represent the carbon and nitrogen atoms, respectively. Defect states are labeled by their irreducible representation and localization. (a) $\text{a}_{1}(\text{N})$, (b) $\text{b}_{2}(\text{N})$, (c) $\text{a}_{1}(\text{C})$, (d) $\text{b}_{1}(\text{C})$, (e) defect levels.}
\label{fig:wavefunc}
\end{figure}

Next, we discuss the nature of electronic structure of the neutral $\text{N}_{2}\text{V}$ and methods to calculate it properly. The four dangling bonds of the defect under $\text{C}_{2\text{v}}$ symmetry produces  $\text{b}_{2}$, $\text{a}_{1}$ and $\text{b}_{1}$ Kohn-Sham levels in the gap, in ascending energy order (see Fig.~\ref{fig:wavefunc}), that may be derived from a split $t_2$ state of the vacancy. In addition, another $\text{a}_{1}$ forms resonant with the valence band. These states are occupied by six electrons. The highest energy occupied in-gap $\text{a}_{1}$ state (HOMO) is a stretched C-C bonding state while the lowest energy empty $\text{b}_{1}$ state (LUMO) is a C-C antibonding state \cite{Zyubin}. As we will show below, HSE06 DFT calculations cannot describe the various multiplet states of the defect caused by the strong correlation of open-shell orbitals. In the following we will use a Hubbard model Hamiltonian to represent the strongly correlating electrons. We particularly focus on the HOMO $\text{a}_{1}$ and LUMO $\text{b}_{1}$ states as active space for the correlated electrons that contribute to the lowest energy excitation configurations.

Our active space with $\text{a}_{1}$ and $\text{b}_{1}$ states may be labeled simply $a$ and $b$, respectively. Then these states may be given as
$a=\frac{1}{\sqrt{2}}\left(A+B\right)$ and $b=\frac{1}{\sqrt{2}}\left(A-B\right)$ where $A$ and $B$ are dangling bonds on the two nearest neighbor carbon atoms. The singlet determinants with $A_1$ symmetry are $\left|^1A_{1(g)}\right>=\left|a^{\uparrow}a^{\downarrow}\right>$, $\left|^1A_{1(e)}\right>=\left|b^{\uparrow}b^{\downarrow}\right>$, $\left|^{1}B_{1}\right>=\frac{1}{\sqrt{2}}\left(\left|a^{\uparrow}b^{\downarrow}\right>-\left|a^{\downarrow}b^{\uparrow}\right>\right)$. The triplet determinants are
$\left|^{3}B_{1}\right>=\left|a^{\uparrow}b^{\uparrow}\right>; \left|a^{\downarrow}b^{\downarrow}\right>$ for $m_{S}=\pm1$, respectively, and $\left|^{3}B_{1}\right>=\frac{1}{\sqrt{2}}\left(\left|a^{\uparrow}b^{\downarrow}\right>+\left|a^{\downarrow}b^{\uparrow}\right>\right)$ for $m_{S}=0$. 
Substituting $a$ and $b$ in the above formulas, the atomic orbital form of the determinants can be obtained. The electronic structure of the neutral $\text{N}_{2}\text{V}$ defect can be described by a Hamilton operator derived from configurational interaction with zero differential overlap (ZDO) approximation and Heisenberg spin coupling. The full Hamiltonian then can be described as a Hamiltonian familiar with the Hubbard model\cite{Hubbard} ($\hat{H}'$) plus the Hamiltonian of the bath of weakly interacting electrons ($\hat{H}_0$)
\begin{align}
\hat{H}=\hat{H}'+\hat{H}_0=\nonumber\\U\left(\hat{n}_{A^{\uparrow}}\hat{n}_{A^{\downarrow}}+\hat{n}_{B^{\uparrow}}\hat{n}_{B^{\downarrow}}\right)\nonumber\\
-\frac{t}{4}\left(\hat{c}^{\dagger}_{A^{\uparrow}}\hat{c}_{B^{\uparrow}}+\hat{c}^{\dagger}_{A^{\downarrow}}\hat{c}_{B^{\downarrow}}+\hat{c}^{\dagger}_{B^{\uparrow}}\hat{c}_{A^{\uparrow}}+\hat{c}^{\dagger}_{B^{\downarrow}}\hat{c}_{A^{\downarrow}}\right)\nonumber\\
+C\left(\hat{n}_{A^{\uparrow}}\hat{n}_{B^{\uparrow}}+\hat{n}_{A^{\downarrow}}\hat{n}_{B^{\downarrow}}+\hat{n}_{A^{\uparrow}}\hat{n}_{B^{\downarrow}}+\hat{n}_{A^{\downarrow}}\hat{n}_{B^{\uparrow}}\right)\nonumber\\
-2J\left({\hat{c}^{\dagger}}_{A^{\uparrow}} {\hat{c}}_{A^{\downarrow}} {\hat{c}}^{\dagger}_{B^{\downarrow}}{\hat{c}_{B^{\uparrow}}}+ {\hat{c}^{\dagger}}_{A^{\downarrow}} {\hat{c}}_{A^{\uparrow}} {\hat{c}}^{\dagger}_{B^{\uparrow}}{\hat{c}_{B^{\downarrow}}}\right)\nonumber\\
-J\left(\hat{n}_{A^{\uparrow}}\hat{n}_{B^{\uparrow}}+\hat{n}_{A^{\downarrow}}\hat{n}_{B^{\downarrow}}-\hat{n}_{A^{\downarrow}}\hat{n}_{B^{\uparrow}}-\hat{n}_{A^{\uparrow}}\hat{n}_{B^{\downarrow}}\right) + \hat{H}_0
\end{align}
where the first term is the onsite repulsion, the second is the hopping, the third is the Coulomb repulsion and the last two terms are from the Heisenberg exchange interaction in the Hubbard Hamiltonian. $\hat{n}$ is the particle number operator while $\hat{c}^{\dagger}$ and $\hat{c}$ is the creation and annihilation operators, respectively. The eigenvalue of $H_0$ is $E_0$ that is the total energy of the bath of weakly interacting electrons that should be added to the solution of $\hat{H}'$. We use the symmetry adapted basis above to write down the matrix of the $\hat{H}'$ Hubbard Hamilton operator,
\begin{equation}
\label{eq:Hmat}
H'=\bordermatrix{~&\left|^1A_{1(g)}\right> &\left|{^1A_{1(e)}}\right>& \left|^{1}B_{1}\right> &\left|^{3}B_{1}\right>\cr ~& \frac{U-t+C+3J}{2}& \frac{U-C-3J}{2}& &\cr &\frac{U-C-3J}{2}&\frac{U+t+C+3J}{2}&&\cr& &&U&\cr & &&&C-J\cr} \text{,}
\end{equation}
where we here neglected the zero-field splitting in the $^{3}B_{1}$ state. In the HSE06 DFT functional calculations the off-diagonal terms are completely neglected, with resulting in $\Psi_1$ and $\Psi_3$ states with $^1A_1$ symmetry.  The $^1B_1$ state is a multideterminant state, and HSE06 DFT cannot calculate the true eigenstate, and as a consequence, the true eigenenergy of the system. Instead, one can calculate 
\begin{equation}\label{eq:psi2}
E\left(\Psi_{2}\right)=\left<a^{\uparrow}b^{\downarrow}\left|H\right|a^{\uparrow}b^{\downarrow}\right>=\frac{U+C-J}{2} + E_0 \text{.}
\end{equation}
Finally, one can calculate the HSE06 DFT total energies for $\Psi_1$, $\Psi_3$, $^3B_1$ (corresponding diagonal terms in Eq.~\ref{eq:Hmat}) and for $\Psi_2$ (Eq.~\ref{eq:psi2}), that provides four equations for the full Hamiltonian parameters. In the full Hamiltonian there are five parameters, however, we are interested in the excitation energies for which three combined parameters, $t$, $J$, and $U-C$, are left (see Eqs.~\ref{ex1A1}-\ref{ex3B1} in the Appendix) that can be derived from the total energy expressions of $\Psi_{1-3}$ and $^3B_1$. 

The HSE06 DFT total energies for these various states can by obtained by $\Delta$SCF calculations \cite{Gali2009}. In order to work with a "non-correlated" basis for $\Delta$SCF energy calculations required in the Hubbard Hamiltonian, we used the basis functions of the closed shell system of the defect in its double negatively charged state $\text{N}_{2}\text{V}$ [Fig. \ref{fig:1}\subref{N2Vstates}], calculated in the $\text{N}_{2}\text{V}^{0}$ groundstate geometry. This choice provides a basis that is free from spin contamination and strong correlation effects.
We note that the relaxed orbitals within unrestricted spinpolarized DFT Kohn-Sham formalism resulting in lower total energies (see Sec.~\ref{sec:app2} in the Appendix). However, our main purpose here is to calculate the excitation energies. As the same restriction on the Kohn-Sham orbitals applies in all the electronic configurations (ground state and excited states), we implicitly assumed that this restriction has the same effect for all the electronic configurations. Finally, the calculated excitation energies are in order.

Other basis that prevents spin contamination could be the neutral state with partially occupied defect levels. However, these basis functions cannot prevent strong correlation effects via Coulomb interaction because of the open shell electronic structure. This is manifested as Kohn-Sham orbitals with broken symmetry which is not a good basis set for a Hubbard calculation. We conclude that the only proper basis is to select the Kohn-Sham wave functions (orbitals) from the closed shell ($2-$) charge state. The total energies in the various occupation of Kohn-Sham states representing the $\Psi_1$, $\Psi_2$, $\Psi_3$, $^3B_1$ multiplets were calculated by keeping these orbitals fixed. We note that this procedure is very different from the usual self-consistent unrestricted spinpolarized DFT method. Consequently, the two approaches result in different excitation energies by $\Delta$SCF method (see Sec~\ref{sec:app2} in the Appendix). Our procedure with fixed orbitals guarantees the proper spin state and symmetry of the single determinant many-body state.
\begin{table}
\caption{HSE06 total energies of considered states of neutral $\text{N}_{2}\text{V}$ relative to that of $\Psi_{1}$ obtained by fixed orbital calculation from the double negatively charged $\text{N}_{2}\text{V}$ basis states in the optimized geometry of the neutral $\text{N}_{2}\text{V}$ by the self-consistent spinpolarized HSE06 calculation. We note that the chosen relative energies correspond to an energy shift of $\frac{t-U-C-3J}{2}$ in $(H')$ (see Eq.~\ref{eq:Hmat}).}
\label{tab:UtCJ}
\begin{ruledtabular}
\begin{tabular}{cc}
state & relative energy (eV)\\
$\Psi_{1}$&$0.00$\\
$\Psi_{2}$&$1.01$\\
$\Psi_{3}$&$2.23$\\
$^{3}B_{1}$&$-0.05$\\
\end{tabular}
\end{ruledtabular}
\end{table}

By calculating the HSE06 DFT energies for the $\Psi_{1-3}$ and $^3B_1$ states (summarized in Table~\ref{tab:UtCJ}), the parameters in the Hubbard Hamiltonian can be calculated \emph{ab initio}, and the resultant values are $t=2.23$~eV, $U-C=2.05$~eV, and $J=0.05$~eV, respectively. The singlet-triplet coupling is minor, and the $U-C$ terms and the hopping term $t$ dominate, $U-C\approx t \approx 2$~eV. A very important finding is that the calculated $\Psi_1\rightarrow\Psi_2$ excitation energy by the self-consistent spinpolarized DFT method scales up by $\approx 1 + \sqrt{2}/2 \approx 1.7$, with respect to the the correct $^1A_{1(g)}\rightarrow {}^1B_1$ excitation energy obtained by the Hubbard model. In other words, the vertical excitation energy associated with $^1B_1$ state increases in the order of eV in the Hubbard model with respect to the HSE06 Kohn-Sham DFT result. As a consequence, the excitation energies of the $^1A_{1(g)}\rightarrow {}^1B_1$ and $^1A_{1(g)}\rightarrow {}^1A_{1(e)}$ transitions are close to each other in the Hubbard model. The error in the self-consistent spinpolarized Kohn-Sham DFT HSE06 functional is much larger than the usual 0.1~eV\cite{deak2014}. Our Hubbard model Hamiltonian derivation clearly shows that the $^1B_1$ state is a particularly highly correlated multiplet which cannot be properly treated by Kohn-Sham hybrid density functionals.  

\section{\emph{Ab initio} magneto-optical parameters}
\label{sec:results}

\begin{figure} 
\centering
\includegraphics[scale=0.9]{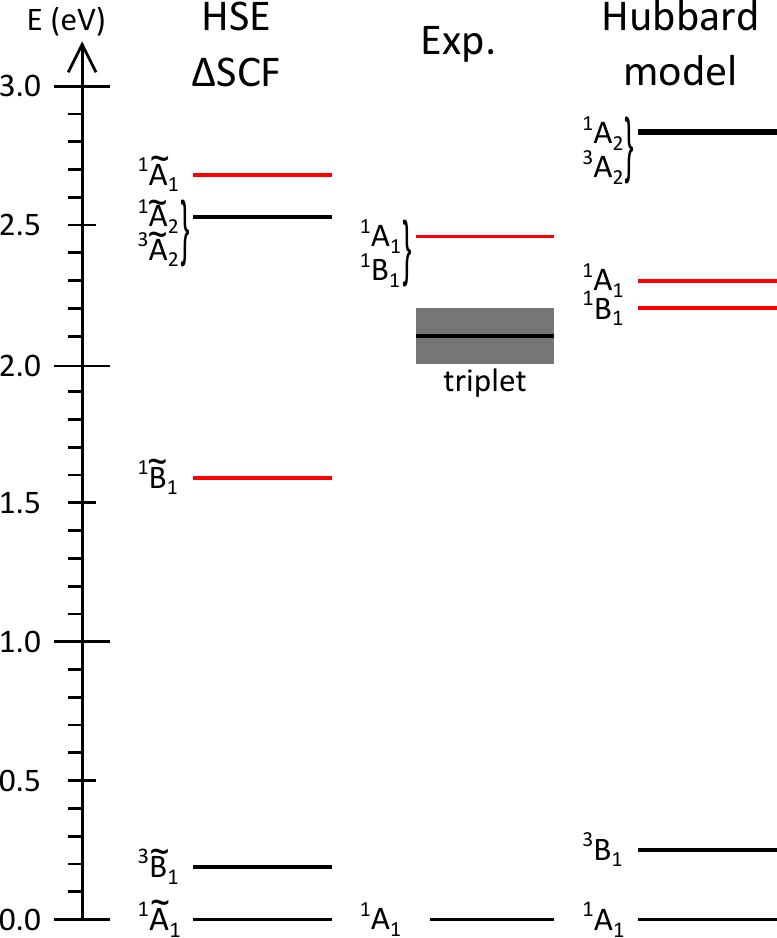}
\caption{Excitation energies of the neutral $\text{N}_{2}\text{V}$ defect in diamond. States that are very close in energy are enclosed with brackets. The lowest optically allowed excitation energies are highlighted with red.  Left panel: self-consistent unrestricted spinpolarized Kohn-Sham HSE06 $\Delta$SCF results. The tilde on the labels of the wavefunctions signs that not the true eigenstate of the system is calculated (see text for more details). Middle panel: experimental zero-phonon-line energies for the singlets (Ref.~\onlinecite{Davies245}) and the proposed energies of triplets (Ref.~\onlinecite{pereira}). Right panel: Hubbard model Hamiltonian results.}
\label{fig:results}
\end{figure}

For direct comparison to the experimental ZPL data, one has to calculate the relaxation energy of ions upon excitation. The relaxation energy was very roughly estimated by self-consistent spinpolarized HSE06 $\Delta$SCF method on 
$\Psi_{1-3}$ states. We find that the relaxation energy on $^1A_{1(e)}$ state is 
$\approx 0.6$~eV whereas it is $\approx 0.2$~eV on $^1B_1$ state. The relaxation energy on $^3B_1$ is small, $0.06$~eV. The final results are depicted in Fig.~\ref{fig:results} that are directly compared to experimental data and the self-consistent spinpolarized Kohn-Sham HSE06 results. Our Hubbard model Hamiltonian with \emph{ab initio} parameters provides significantly improved results over those obtained by the usual self-consistent unrestricted spinpolarized Kohn-Sham HSE06 method. 
We find that the $^1A_{1(g)}\rightarrow {}^1B_1$ ZPL energy is indeed around 2.4~eV, and the $^1A_{1(g)}\rightarrow {}^1A_{1(e)}$ ZPL energy is slightly above that. These are the optically allowed transitions. Higher energy singlet and triplet states with $b_{2}\rightarrow b_{1}$ excitation may form with optically forbidden $A_{2}$ symmetry. The $^1A_{2}$ state cannot absorb light but can play a role in the non-radiative decay when the electron is excited to the H13 band which corresponds to the valence band to $b_{1}$ transition (see Fig.~\ref{fig:results}). The total energy of the $^{3}A_{2}$ could be calculated $\Delta$SCF procedure from  ${^{3}B_{1}}\rightarrow {^{3}A_{2}}$ excitation energy whereas the total energy of $^{1}A_{2}$ should be slightly higher due to the small singlet-triplet coupling $J$. 

Regarding the triplet energy levels, their energies in the region of $270-480$~meV below the ${^{1}B_{1}}$ level were proposed from PL lifetime measurements where they found a delayed luminescence of millisecond lifetime\cite{pereira} that they associated with a spin-orbit mediated tunneling process\cite{freed1970} from the metastable triplets to the lowest energy singlet excited state with the formula
\begin{equation}
W(T)=\frac{K}{\sqrt{kT^{*}}}\coth\left(\frac{\hbar\omega}{2kT}\right)\exp\left(-\frac{E_a}{kT^{*}}\right)
\end{equation}
\begin{equation}
K=\frac{\left|C_{sl}\right|^2\omega\sqrt{2\pi}}{\sqrt{2E_{M}}}
\end{equation}
\begin{equation}
C_{sl}=\frac{\left<s\left|\hat{H}_\text{SO}\right|l\right>J_{sl}}{E_s-E_l} \text{,}
\end{equation}
with $kT^{*}=\frac{1}{2}\hbar\omega\coth\left(\hbar\omega/2kT\right)$, $E_a$ is the barrier energy between the corresponding states, $\hbar\omega$ dominant phonon frequency, $E_M$ is the relaxation energy between the two states, $k$ is the Boltzmann constant and $T$ is the temperature in Kelvin. $C_{sl}$ is the coupling of states $s$ and $l$ where $\hat{H}_\text{SO}$ is the spin-orbit coupling operator, $J_{sl}$ is the electron-phonon coupling. We found that the $^3B_1$ level is rather far ($> 2$~eV) from the excited singlet states, thus we estimated its delayed luminescence. As the transition from ${^{3}B_{1}}$ to ${^{1}B_{1}}$ is a second order process which should be presumably slow, we calculated the transition to ${^{1}A_{1}}$ with first order ISC process. The estimated strength of spin-orbit coupling is around $10$~GHz and that of the electron-phonon coupling is $0.1~\sqrt{\text{eV}}$, similar to the values in NV center\cite{Batalov2009, Thiering2017}. From our vibrational analysis, calculated with the computationally less expensive PBE functional\cite{PBE} using PBE optimized geometries, we obtained the average phonon energy of $84$~meV weighted by the partial Huang-Rhys factors\cite{Huang} for the vibrational coupling for this transition. The relaxation energy between ${^{1}A_{1}}$ and ${^{3}B_{1}}$ is estimated to be $0.4$~eV. With these parameters we obtain a lifetime in the order of $10^6$~s at various temperatures. This implies that the observed delayed luminescence is not intrinsic to the defect. As the delayed luminescence was observed for the ensemble of N$_2$V defects we speculate that it originates from the interaction with other defects in diamond. Future single defect measurements may conclude the nature of this emission.

\begin{figure}[H]
\centering
\includegraphics[scale=0.14]{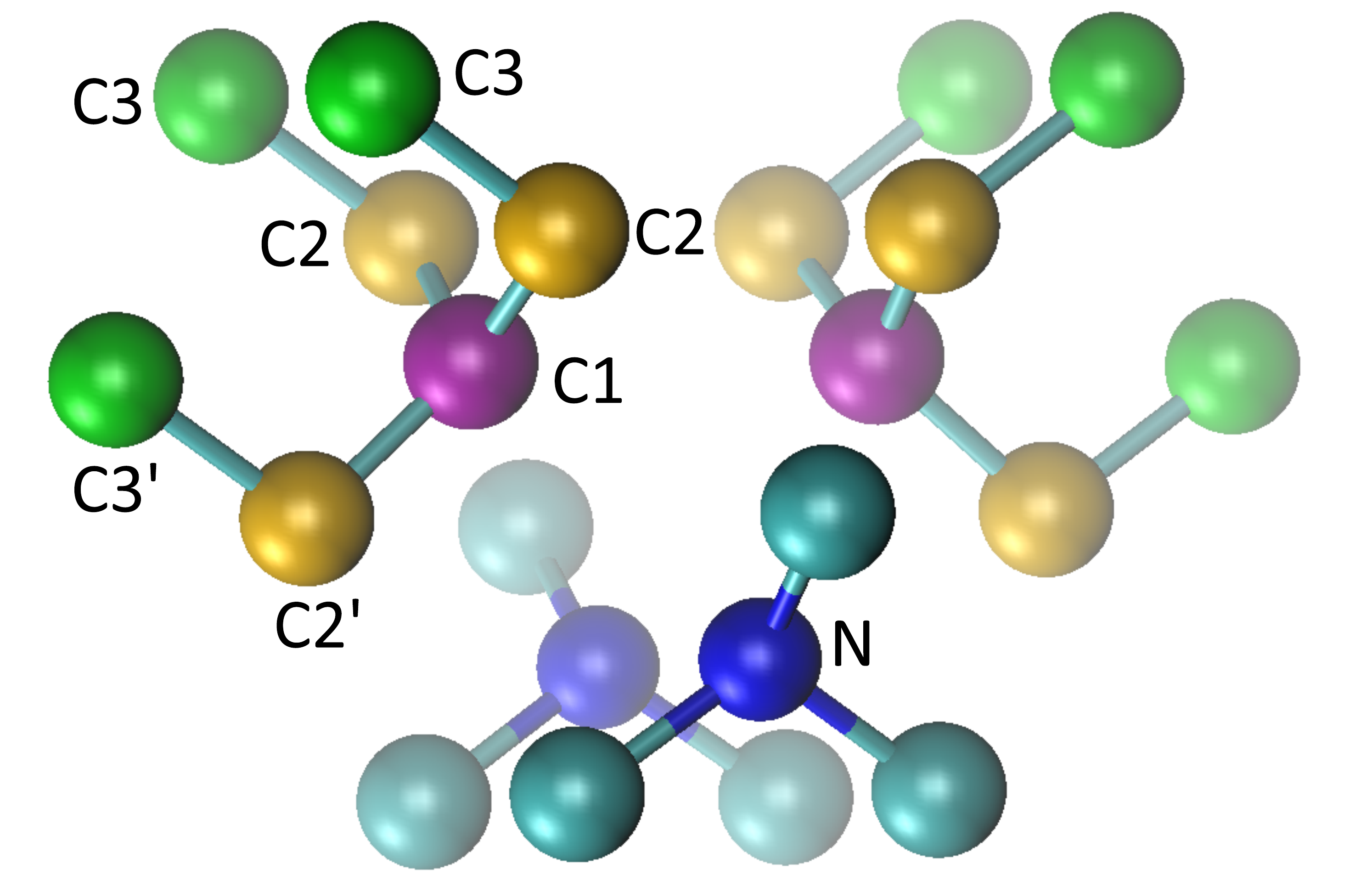}
\caption{Visualization of atomic sites with dominant hyperfine parameters shown in Table~\ref{sup:hyperfine} for $\text{N}_{2}\text{V}^{0}$.}
\label{fig:hyperfine}
\end{figure}

In the following, we present the calculated hyperfine interaction of the electron spin with $^{14}$N and proximate $^{13}$C nuclear spins. The identification of $^{13}\text{C}$ sites with dominant hyperfine parameters in the vicinity of $\text{N}_{2}\text{V}$ defect is of great importance in quantum memory realization with this defect as they can store the quantum information for relatively long time. These sites are highlighted in Fig. \ref{fig:hyperfine} and the corresponding calculated hyperfine parameters are listed in Table \ref{sup:hyperfine}.
\begin{table}[H]
\caption{HSE DFT calculated hyperfine constants for $\text{N}_{2}\text{V}^{0}$ for $^{14}$N and $^{13}$C nuclei. The location of $^{13}$C nuclei is depicted in Fig.~\ref{fig:hyperfine}.}
\label{sup:hyperfine}
\begin{ruledtabular}
\begin{tabular}{cccc}
site&$A_{xx}$ (MHz)&$A_{yy}$ (MHz)&$A_{zz}$ (MHz)\\
$^{14}\text{N}$&$9.6$&$9.5$&$14.5$\\
C1&$82.3$&$81.7$&$198.9$\\
C2&$-11.5$&$-8.5$&$-12.1$\\
C2'&$-11.2$&$-8.3$&$-11.6$\\
C3&$15.9$&$15.7$&$24.2$\\
C3'&$16.1$&$15.9$&$24.4$\\
\end{tabular}
\end{ruledtabular}
\end{table}

\section{Summary}

We analyzed the neutral $\text{N}_{2}\text{V}$ defect in diamond by means of \emph{ab initio} calculations, and concluded that a quantum memory can be realized by this defect controlled by optical excitation and microwave manipulation. We showed that the electronic structure of this defect is a prototype of highly correlated states that can be treated by our method that is a combination of density functional theory and a Hubbard model.

\section*{Acknowledgment}
We acknowledge the support from EU Commission on the DIADEMS project (contract No.~611143).

\appendix
\section{Additional formulas for the Hubbard model}
\label{sec:append}
In this section, we explicitly show the connection between the DFT total energies of the single determinant states and the multiplets, and write down the formulas that we applied to obtain the Hubbard parameters and calculate the excitation energies.

By using the definitions of $a$ and $b$ orbitals in the main text, one can express the corresponding wavefunctions in terms of $A$ and $B$ dangling bonds as
\begin{align}\label{AO1}
\left|^1A_{1(g)}\right>&=\frac{1}{2}\left(\left|A^{\uparrow}A^{\downarrow}     \right>+\left|B^{\uparrow}B^{\downarrow}\right>\right)+\frac{1}{2}\left(\left|B^{\uparrow}A^{\downarrow}\right>-\left|B^{\downarrow}A^{\uparrow}\right>\right) \\\label{AO2}
\left|^1A_{1(e)}\right>&=\frac{1}{2}\left(\left|A^{\uparrow}A^{\downarrow}\right>+\left|B^{\uparrow}B^{\downarrow}\right>\right)-\frac{1}{2}\left(\left|B^{\uparrow}A^{\downarrow}\right>-\left|B^{\downarrow}A^{\uparrow}\right>\right) \\\label{AO3}
\left|^{1}B_{1}\right>&=\frac{1}{\sqrt{2}}\left(\left|A^{\uparrow}A^{\downarrow}\right>-\left|B^{\uparrow}B^{\downarrow}\right>\right) \\\label{AO4}\left|^{3}B_{1}\right>&=\left\{\begin{array}{lr}\left|A^{\uparrow}B^{\uparrow}\right>\\
 \frac{1}{\sqrt{2}}\left(\left|B^{\uparrow}A^{\downarrow}\right>+\left|B^{\downarrow}A^{\uparrow}\right>\right)\\
 \left|A^{\downarrow}B^{\downarrow}\right>\end{array}\right..
\end{align}

By applying the Hubbard Hamiltonian in the main text, one obtains Hamiltonian matrix in the basis of $\left|^1A_{1(g)}\right>$, $\left|^1A_{1(e)}\right>$, $\left|^{1}B_{1}\right>$, $\left|^{3}B_{1}\right>$ as shown in the main text. After diagonalization of the Hubbard Hamiltonian, the resultant vertical excitation energies are
\begin{equation}\label{ex1A1}
E\left(^1A_{1(g)}\rightarrow {^1A_{1(e)}}\right)=\sqrt{t^{2}+\left(U-C-3J\right)^{2}}.
\end{equation}
\begin{equation}\label{ex1B1}
E\left(^1A_{1(g)}\rightarrow {^{1}B_{1}}\right)=\frac{U-C-3J+\sqrt{t^{2}+\left(U-C-3J\right)^{2}}}{2}
\end{equation}
\begin{equation}\label{ex3B1}
E\left(^1A_{1(g)}\rightarrow {^{3}B_{1}}\right)=\frac{C-U-5J+\sqrt{t^{2}+\left(U-C-3J\right)^{2}}}{2}.
\end{equation}

\section{Additional information about the raw DFT total energies}
\label{sec:app2}
We show in Table \ref{tab:scfHub} that calculation of the total energies in the Hubbard model from unrestricted spinpolarized Kohn-Sham DFT HSE06 orbitals is not appropriate.

The total energy of $^{1}\tilde{A}_{1(g)}$ relative to that of $\Psi_{1}$ is $-6.75$~eV as listed in Table~\ref{tab:Ax1} because we applied restriction to the Kohn-Sham orbitals as explained in the main text. The calculated excitation energies within the Hubbard model taking the values in Table~\ref{tab:Ax1} are given in Table~\ref{tab:scfHub}. These results are very far from the experimental data. This can be understood by considering the fact that the unrestricted spinpolarized HSE06 Kohn-Sham orbitals are spin contaminated and \emph{break} the symmetry of the system. Thus, these orbitals are not suitable for serving as basis for Hubbard model as Hubbard model requires orbitals with appropriate symmetry and spin state. 
\begin{table}
\caption{HSE06 total energies of considered states of neutral $\text{N}_{2}\text{V}$ relative to that of  $^{1}\tilde{A}_{1(g)}$ are obtained by self-consistent unrestricted spinpolarized HSE06 calculation in the optimized geometry of the neutral $\text{N}_{2}\text{V}$. The tilde over the electronic states labels that those states are not the true symmetrical eigenstates of the system.}
\label{tab:Ax1}
\begin{ruledtabular}
\begin{tabular}{cc}
state & relative energy (eV) \\ \hline
$^{1}\tilde{A}_{1(g)}$&0.00\\
$^{1}\tilde{B}_{1}$&$1.67$\\
$^{1}\tilde{A}_{1(e)}$&$2.92$\\
$^{3}\tilde{B}_{1}$&$0.25$\\
\end{tabular}
\end{ruledtabular}
\end{table}
\begin{table}
\caption{Vertical excitation energies in the Hubbard model calculated from self-consistent unrestricted spinpolarized HSE06 Kohn-Sham orbitals}
\begin{ruledtabular}
\begin{tabular}{cc}
excitation&vertical excitation energy (eV)\\ \hline
$^{1}A_{1(g)}\rightarrow {^{3}B_{1}}$&$1.07$\\
$^{1}A_{1(g)}\rightarrow {^{1}B_{1}}$&$3.91$\\
$^{1}A_{1(g)}\rightarrow {^{1}A_{1(e)}}$&$4.39$
\end{tabular}
\end{ruledtabular}\label{tab:scfHub}
\end{table}


%

\end{document}